\documentclass[journal]{IEEEtran}

\ifCLASSINFOpdf
\else
   \usepackage[dvips]{graphicx}
\fi
\usepackage{url}
\usepackage{amsmath}
\usepackage{graphicx}
\usepackage{placeins}
\usepackage{float}
\usepackage{bbm}
\usepackage{subfigure}
\usepackage{verbatim}
\usepackage{amssymb}
\usepackage{amsmath}
\usepackage{amsthm}
\usepackage{mwe}
\usepackage{docmute}
\usepackage[ruled, vlined]{algorithm2e}

\usepackage[disable]{todonotes}

\usepackage{graphicx}

\newtheorem{definition}{Definition}

\begin{document}

\title{Decentralized Bandits with Feedback for Cognitive Radar Networks}
\author{William W. Howard, R. Michael Buehrer
\thanks{W.W. Howard and R.M. Buehrer are with Wireless@VT, Bradley Department of ECE, Virginia Tech, Blacksburg, VA, 24061. \\
e-mails:$\{$wwhoward, buehrer$\}$@vt.edu  }
}

\maketitle
\pagenumbering{roman}

\begin{abstract}
    Completely decentralized Multi-Player Bandit models have demonstrated high localization accuracy at the cost of long convergence times in cognitive radar networks. 
    Rather than model each radar node as an independent learner, entirely unable to swap information with other nodes in a network, in this work we construct a `` central coordinator'' to facilitate the exchange of information between radar nodes. 
    We show that in interference-limited spectrum, where the signal to interference plus noise (\texttt{SINR}) ratio for the available bands may vary by location, a cognitive radar network (CRN) is able to use information from a central coordinator to reduce the number of time steps required to attain a given localization error. 
    Importantly, each node is still able to learn separately. 
    We provide a description of a network which has hybrid cognition in both a central coordinator and in each of the cognitive radar nodes, and examine the online machine learning algorithms which can be implemented in this structure. 
    
\end{abstract}

\section{Introduction}
We consider a problem where a group of cognitive radar nodes must cooperate to localize a target, while the qualities of each radar observation vary by radar node and over time. 
Coexistence in radar is an increasingly important problem to study, as communication bands begin to occupy frequencies which were previously dedicated to radar use. 
We will describe a low-power, distributed radar system capable of cooperatively localizing a target while autonomously optimizing channel selections to increase \texttt{SINR}.
In our scenario, the \texttt{SINR} varies with time due to interference behavior as well as the relative target motion.

Distributed radar provides many benefits over single-node approaches \cite{5703085}. 
Namely, distributed radar provides spatial diversity and improved resilience to outages. 
Cognitive techniques, on the other hand, provide a radar system the capability to modify operating parameters to improve performance \cite{haykin2006}. 
CRNs, then, are capable of greater performance due to increased diversity as well as greater flexibility due to cognitive strategies. 
The CRN we will investigate is distributed throughout space and has the ability to monitor several different channels. 
Each channel provides a different \texttt{SINR}, and each radar node in the network must pick one channel in each Coherent Pulse Interval (CPI). 
Multiple radar nodes cannot select the same channel at the same time, so as to avoid interfering with each other.

In one of the original works on cognitive radar networks \cite{1574168}, Haykin defined two different cognition modes: 
\begin{enumerate}
    \item \textbf{Distributed Cognition}, where observations from the individual nodes are combined at a \emph{fusion center} but no feedback is provided to the nodes. 
    \item \textbf{Centralized Cognition}, where a \emph{central coordinator} is the only cognitive agent, collecting observations from each node and dictating future actions. 
\end{enumerate}
Implementations of centralized cognition are susceptible to failure of the central coordinator, and also must deal with a large communication overhead due to the information that must be exchanged. 
On the other hand, fully distributed cognition approaches lack flexibility due to the need to coordinate actions in a decentralized manner. 
This leads to long convergence times, which are damaging when a target is only available for a brief time. 
In this paper, we will describe a hybrid approach, where the radar nodes and central coordinator each exert some control over the network's actions. 
In this hybrid scheme, the central coordinator acts as a fusion center to combine radar measurements. 
The central coordinator also possesses cognitive functions, capable of communicating reward weights observed throughout the network to each node.

\paragraph{Contributions} 
In this work we develop a method to make decisions in a CRN in a centralized manner, while preserving the independence of each node. 
We show that in a non-stationary environment, where the observed rewards may change over time, a central coordinator is able to adjust the rate at which feedback is provided to maximize tracking performance while minimizing the incurred communication overhead. 
Since the network is centralized, decisions can be made in a manner which eliminates mutual interference, separating this work from decentralized techniques. 
Specifically, we contribute the following to the state of the art: 
\begin{itemize}
    \item The first online machine learning solution for sequential action selection in radar networks where channel rewards vary over time. 
    \item The first description of a hybrid cognition model, where the central coordinator performs data fusion \emph{as well as} cognitive functions to improve the performance of the radar network. 
\end{itemize}

\paragraph{Notation} We use the following notation. 
Matrices and vectors are denoted as bold upper $\mathbf{X}$ or lower $\mathbf{x}$ case letters.
Functions are shown as plain letters $F$ or $f$. 
Sets $\mathcal{A}$ are shown as script letters. 
The cardinality $|\mathcal{A}|$ of a set $\mathcal{A}$ refers to the number of elements in that set. 
The transpose operation is $\mathbf{X}^T$. 
The set of all real numbers is $\mathbb{R}$ and the set of integers is $\mathbb{Z}$. 
The speed of electromagnetic radiation in a vacuum is given as $c$. 
Element-wise (scalar) multiplication is denoted as $x*y$. 
The Euclidean norm of a vector $\mathbf{x}$ is written as $||\mathbf{x}||$. 

\paragraph{Organization} 
Section II discusses prior work in this area, and frames our contribution in the literature. 
Section III then introduces our specific network structure and discusses the mathematical models we use. 
Section IV discusses the learning problem that we consider and provides details on the applicable online machine learning algorithms. 
In Section V we provide simulations to support our work, and in Section VI we draw conclusions and provide discussion on future directions.



\section{Background}
Cognitive radar has been the subject of much study in recent years. 
In \cite{Martone_CRN_loop}, the authors give details on recent work in cognitive radar for spectrum sharing. 
Since cognitive radar is inherently flexible, it is easily applied to spectrum sharing problems where radar systems are secondary users.

The authors of \cite{9455255} and \cite{thornton2022_universaljournal} apply detailed machine learning techniques to single-node cognitive radar, describing waveform selection techniques which adapt to a broad class of target categories.

Cognitive radar networks have also been of interest in the literature. 
The authors of \cite{howard2021_multiplayerconf} and \cite{howard2022_MMABjournal} provided the first description of multi-player bandit algorithms applied to the CRN problem. 
In their approach, the reward for a given action was assumed to be constant between nodes and over time. 
Our current work differs in that the observed \texttt{SINR} varies by both node and time.

In \cite{howard2022_adversarialconf}, the authors investigated the CRN problem where node rewards are selected by an adversary. 
This models the scenario where an interferer attempts to force the CRN to experience poor performance. 
Our current work considers the case where interferers are oblivious to the CRN and have constant behavior over time.

Distributed learning is a broad class of machine learning approaches. 
Generally, distributed learning problems focus on a group of nodes which seek performance improvements which scale with group size. 
Since the CRN channel selection problem is also a sequential learning problem, we draw from the multi-player multi-armed bandit literature. 
Multi-armed bandits (MABs) are game theoretic models which consider a set of actions which each offer some reward. 
Over finite time steps, MAB algorithms describe the sequence of actions a learner should select to maximize the cumulative reward \cite{bandits}. 
Multi-player MAB models extend this to several identical players \cite{pmlr-v48-rosenski16}.

Federated learning is a separate ML structure where instead of sharing data to collaborate, nodes in a network will train separate models to preserve privacy between data sets. 
This model differs from our approach because federated learning assumes data is independent and identically distributed between node data sets. 
Federated learning has been applied to sensor networks \cite{8950073}, where models trained at various nodes are combined to form a single model. 
Our technique focuses on a single model learning from various data sources.

\subsection{Problem Summary}
As mentioned above, the problem where \texttt{SINR} is constant over time and node has been studied \cite{howard2022_MMABjournal}. 
However, we will consider the case where \texttt{SINR} varies over time and by node. 
This is caused by the motion of the target. 
As described in the radar equation \cite{handbook}, the power of received pulses is dependent on the fourth power of the distance between transceiver and target. 
Due to this relationship, as the target moves through the environment, different nodes will experience different \texttt{SINR} at different times. 
We will describe a method to solve this problem, and to predict which actions will be best given the motion of the target.

Before a CRN is able to determine the optimal actions to select, it must first learn about the environment. 
In our problem, this will consist of estimating the \texttt{INR} in each channel. 
So, we need to use an algorithm capable of 1) avoiding collisions, and 2) exploring each channel enough times to develop an estimate of the \texttt{SINR}, which varies slowly with time. 
This information, coupled with target position estimates provided by the central coordinator, is enough to estimate the \texttt{INR} of each channel. 
Note that we assume the channels are identical at each radar node. 
If the target were equidistant from each node, we would expect identical observation quality at each node. 
\texttt{SINR} varies by radar node due to the target position.

In an environment with a single target of interest, the accuracy of each radar measurement will vary with the inverse of SINR \cite{radar_channel_quality}. 
In other words, as the spectrum becomes more congested, radar nodes will experience a reduction in performance. 
When multiple channels with differing interference statistics are available to a radar node, we are presented with a learning and optimization problem. 
In the case where a single node is considered, that node can try each channel multiple times and determine which will be best. 
The problem is similar when multiple nodes are considered, with the important difference that some effort is required to prevent multiple nodes selecting the same channel\footnote{This would increase the interference in that channel and reduce the \texttt{SINR} observed by each node. }.
Since we are considering a centralized problem, these instances of \emph{collisions} are avoidable through coordination.

From the above description, it is clear that the radar nodes will experience different quality observations as the target moves through the environment. 
In addition, in order to determine which actions will be optimal, the CRN will need to learn 1) the interference in each channel, and 2) the position of the target over time. 
This structure is convenient, since the overall goal of the network is target localization.

\section{Network Structure}
The general structure of our network is as follows. 
The $M$ radar nodes are distributed uniformly at random throughout the square kilometer. 
We denote the position of each node $R_m$ ($m\leq M$) as $\mathbf{x}_m = [x_m, y_m]$. 
Since the central coordinator can provide feedback, these positions are known to the radar nodes. 
Call $\mathcal{M}$ the set of all radar nodes. 
The scene contains one target. 
Denote the position of the target at a time $t$ as $\mathbf{x}(t) = [x(t), y(t)]$. 
Each radar node is able to transmit Linear Frequency Modulated (LFM) chirp waveforms. 
The set $\mathcal{C}$ of channels available to the radar nodes consists of $N$ orthogonal channels of equal bandwidth.

The CRN divides time into CPIs, and further into Pulse Repetition Intervals (PRIs). 
Each CPI consists of 1000 PRIs. 
In each CPI, each node $R_m \in \mathcal{M}$ executes the following steps: 
\begin{enumerate}
    \item Select a channel $C_n\in \mathcal{N}$ from the available channels, and transmit a train of 1000 LFM pulses. 
    \item Receive the waveform and process to determine range $\hat{r}_m(t)$, velocity $\hat{\dot{r}}_m(t)$ and angle $\hat{\theta}_m(t)$ estimates. 
    \item Transmit the target state estimates to the coordinator. 
    \item Receive a position estimate from the coordinator. 
    \item Update target tracking filter. 
    \item Determine the channel metric $P_c$ for the selected channel (defined later). 
    \item Update channel selection algorithm, requesting feedback from the central coordinator as necessary. 
\end{enumerate}
Meanwhile, the central coordinator performs the following functions: 
\begin{enumerate}
    \item Receive target state estimates. 
    \item Process these to determine a target position estimate. 
    \item Transmit the position estimate to the radar nodes. 
    \item Provide channel selection feedback in the form of action sequences using a cognitive strategy. 
\end{enumerate}

Importantly, we assume that the CRN consists of low cost, low-complexity nodes. 
This has several important implications. 
\begin{itemize}
    \item In order to conserve power, the radar nodes only conduct signal processing once per CPI. 
    This means that any channel estimation must be conducted while radar returns are expected to be present. 
    \item The transmit arrays of each radar node have sufficient gain to illuminate the target and are electronically steerable. 
\end{itemize}
The central coordinator will require the use of a localization algorithm to use the noisy range estimates to produce a position estimate. 

\subsection{Target and Channel Modeling}
The target platform is assumed to be an aircraft with a velocity of $200\texttt{m/s}$. 
We assume a point target with a constant RCS. 
Further, let the scenario take place in an area with sufficient clutter that interfering sources do not have a strong directional dependence. 
So, each radar node will observe similar interference and noise power in each channel but different \texttt{SINR}.

In order to estimate the interference and noise power of the channel, the radar nodes need to establish an estimate of the target range. 
Assuming the target has an RCS which is constant over frequency and angle, we can rearrange the radar range equation as follows. 
\begin{equation}
    P^* = \frac{P_r}{\sigma} = \frac{P_tG^2\lambda^2}{(4\pi)^3\hat{r}_m^4}
\end{equation}
Once the range estimate $\hat{r}_m$ is determined, the radar node can estimate the power returned in each pulse from the target. 
We will define the channel metric as: 
\begin{equation}
    P_c = \texttt{SINR}_{dB} - P^*_{dB}
\end{equation}
We assume that the local variations are sufficient to provide different values at each radar node, but not so much as to cause the \emph{order} of these values to change. 
So, if one node experiences $P_c(a) > P_c(b)$ for two channels $C_a, C_b$, all nodes will observe the same. 
We assume that each radar node has the capability to measure \texttt{SINR}. 

In the long term, due to repeated observations, each radar node will converge to the true values of $P_c$ for each channel. 
Denote these values in the row vector $\mathbf{P_m} = [P_c(1), P_c(2), \dots, P_c(N)]$ for each node $R_m$. 
With knowledge of each node's position and the target position, a vector of ranges $\overline{\mathbf{r}}$ can be developed, where $\overline{\mathbf{r}}(m) = ||\mathbf{x}_m - \hat{\mathbf{x}}(t)||$

\subsection{Target Localization}


Each radar node $R_m$ processes estimates for range $\hat{r}_m$, velocity $\hat{\dot{r}}_m(t)$ and angle $\hat{\theta}_m(t)$ estimates.
These are processed at each node to determine a position estimate $\hat{\mathbf{x}}_m$. 
The measurements are combined at the central coordinator and used to update a Kalman filter, producing a final position estimate $\hat{\mathbf{x}}$.

\section{Candidate Algorithms}
\subsection{Matchings and Utility}
Through the selection of a weight matrix, we can identify the optimum assignment from radar nodes to channels. 
A weight matrix $W(t)$ is the $M$ by $N$ matrix consisting of the reward a radar node $R_m$ will observe for selecting a channel $C_n$ at time $t$.  
Algorithms must select from the set of all matchings $\Pi$, where matchings are defined below. 

\begin{definition}[Matching]
    A matching $\pi(t):\mathcal{M} \to \mathcal{N}$ is any assignment from the set of radar nodes $\mathcal{M}$ to the set of channels $\mathcal{N}$ with no common vertices. In other words, matching functions $\pi \in \Pi$ are injective. 
\end{definition}

Obviously, since the weights vary by radar node - channel pairing, some matchings will be better than others. 

\begin{definition}[Utility]
    The \emph{utility} of a mapping $\pi$ is the sum of the rewards each radar observes using that mapping: 
\begin{equation}
    \label{eq:matching_utility}
    U(\pi) = \sum_{R_m C_n \in \pi} W_{m,n}
\end{equation}
\end{definition}

Further, there will be one or more matching with greatest utility. 

\begin{definition}[Optimal Matching]
    If a matching $\pi(t) \in \Pi$ has maximum utility, $U(\pi(t)) = U^*(t)$, it is called optimal and denoted $\pi^*(t)$. 
    \begin{equation}
        \label{eq:max_utility}
        U^*(t) = \max_{\pi \in \Pi} U(\pi(t))
    \end{equation}
\end{definition}

The goal of any selection algorithm is to select an optimal matching, $\pi^*(t)$. 
When an algorithm selects $\pi(t) \neq \pi^*(t)$, we can measure the \emph{regret} \cite{bandits} as the difference in utility. 

\begin{definition}[Regret]
    The regret for an algorithm at time $t$ is the difference in utility between the selected matching $\pi(t)$ and the optimal matching $\pi^*(t)$. 
    \begin{equation}
        \rho(t) = U(\pi^*(t)) - U(\pi(t))
    \end{equation}
\end{definition}

Note that the regret in each time step will be positive, since the utility will always be less than or equal to the optimal utility. 
We can then define the cumulative regret as the sum of all regret until time step T: 
\begin{equation}
    \rho_T = \sum_{t=1}^T \rho(t)
\end{equation}

\subsection{Rewards}
We will define two different sets of rewards. 
The first, $\mathbf{S}$, captures the observed \texttt{SINR} by each radar node for each channel. 
The second separates the two underlying processes to form estimates of both $\mathbf{P}_c$ and $\hat{\mathbf{r}}$

So, using our assumptions that the channel interference varies by location but not enough to change their ranking, 

Once estimates of the channel and target states are formed, each node can determine a matrix of weights, which represents the observation quality for each node-action pairing. 
Since we're attempting to optimize the network average \texttt{SINR}, we can form this matrix as

\begin{equation}
    \mathbf{W} = \frac{1}{\overline{\mathbf{r}}^T} *
    \begin{bmatrix}
        \mathbf{P}_1 \\
        \mathbf{P}_2 \\
        \vdots \\
        \mathbf{P}_M
    \end{bmatrix} = \frac{1}{\overline{\mathbf{r}}^T} * \overline{\mathbf{P}}
\end{equation}

Note that the multiplication here is element-wise, and recall that $\overline{\mathbf{r}}(m) = ||\mathbf{x}_m - \hat{\mathbf{x}}(t)||$. 
We specify multiplication here instead of addition for two reasons. 
First, if we were to use addition, the maximum utility matching for these weights would not change between arbitrary $\overline{\mathbf{r}}$. 
In addition, we can choose to optimize the closest-range node's observations, which is accomplished through the multiplication. 
Note that for either scenario (multiplication or addition), the optimal matching will consist of the same channels, just with a different ordering. 

\subsection{Algorithms}

\subsubsection{Oracle}
An oracle for this problem knows the true channel metrics and can predict ranges perfectly. 
Therefore, in each CPI $t$, the oracle will select $\pi^*(t)$. 
Specifically each node $R_m$ selects channel $\pi^*(t)_m$. 
The regret for this oracle will be 0. 

\subsubsection{Centralized Multiplayer Explore-Then-Commit}
ETC \cite{mehrabian20a} uses the Upper Confidence Bound (UCB) \cite{UCB_fischer} as a threshold to narrow down the space of possible matchings $\Pi$ over time. 
Since each node has a ``rank'', they can play through a given sequence of matchings to observe the \texttt{SINR} in each channel sequentially. 
Over time, the algorithm identifies which matchings have utilities that are below a threshold, and eliminates them from consideration. 
We make the modification that the central coordinator is responsible for processing the UCB values, and informing the nodes of the rewards observed throughout the network.

ETC establishes a sequence $\Gamma$ of matchings $\pi$. 
The sequence is determined by the central coordinator, and once it is exhausted a new one is provided. 
Once the sequence is a single matching, the algorithm enters an \emph{exploitation} phase, where it seeks to minimize regret by selecting the action it has determined to be best. 
Let $|\Gamma|$ denote the number of matchings in $\Gamma$. 
Recall that $\mathbf{S}$ denotes the $\texttt{SINR}_{dB}$ reward matrix, and denote the optimal matching for $\mathbf{S}$ as $\pi^*(\mathbf{S})$. 
Algorithm \ref{algo:ETC} provides the node-level cognition for this algorithm.

\vspace{0.1in}
\begin{algorithm}[h!]
\SetAlgoLined
\KwResult{C(t)}
 \uIf{$|\Gamma| != 1$}{
  $C(t) = \Gamma(t)_m$\;
  }
 \Else{
  $C(t) = \pi^*(\mathbf{S})_m$\;
  }
 \caption{Explore-Then-Commit for node $R_m$}
 \label{algo:ETC}
\end{algorithm}
\vspace{0.1in}

Once the sequence $\Gamma$ is explored, the central coordinator provides a refined sequence, removing any matchings below a threshold determined by UCB. 
The matchings in the $p^{th}$ sequence are explored $2^p$ times each, which causes each successive sequence to require more rounds to explore. 
The required feedback communication overhead decreases with time. 

\subsubsection{Centralized Multiplayer Explore-Then-Predict}
We make a slight modification to CM-ETC by specifying that once the algorithm has converged, we begin using $\mathbf{W}$ instead of $\texttt{SINR}_{dB}$. 
This is because while we need to use $\texttt{SINR}_{dB}$ while the algorithm is learning to avoid errors, once it has converged we should expect good estimates of both $\overline{\mathbf{P}}$ and of $\overline{\mathbf{r}}$. 
According to the oracle, however, we should expect higher \texttt{SINR} using the more time-dependent weights in $\mathbf{W}$.

Similarly to ETC, ETP relies on a central coordinator specified sequence of matchings to dictate the exploration phase. 
Then, in the exploitation phase, ETP uses the time-dependent weights in $\mathbf{W}$ to determine actions. 
Since each node will converge to the same conclusion and shares the same target location information, the action sequences will remain synchronized. 

\vspace{0.1in}
\begin{algorithm}[h!]
\SetAlgoLined
\KwResult{C(t)}
 \uIf{$|\Gamma| != 1$}{
  $C(t) = \Gamma(t)_m$\;
  }
 \Else{
  $C(t) = \pi^*(\mathbf{W})_m$\;
  }
 \caption{Explore-Then-Commit for node $R_m$}
 \label{algo:ETP}
\end{algorithm}
\vspace{0.1in}

\subsubsection{Random Matchings}
This naive algorithm has the radar nodes select a random matching $\pi(t)$ from $\Pi$ in each time step.


\section{Results}
\begin{figure}
    \centering
    \includegraphics[scale=0.65]{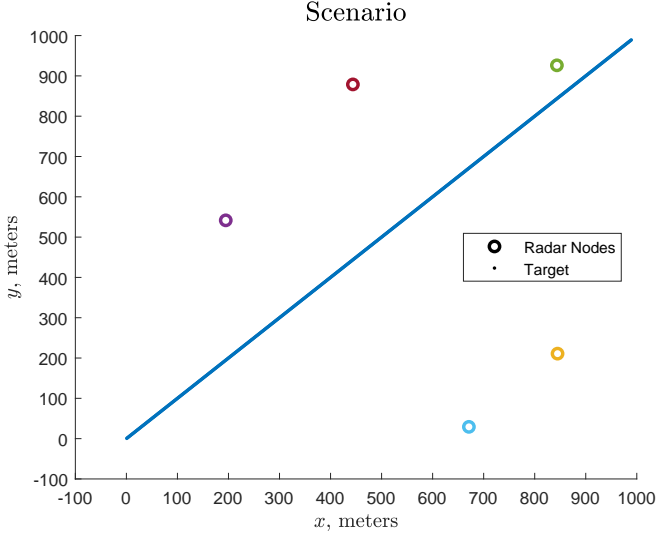}
    \caption{Target scenario. The radar nodes are located randomly and must collaborate to localize the target in each CPI. The target is initially at the origin and moves towards $[1,1]$\texttt{km} at $200$\texttt{m/s}. }
    \label{fig:Scene}
\end{figure}

Our simulations consist of five radar nodes and one target. 
The radar nodes are randomly placed in the unit kilometer. 
The target is initially at $[0,0]\texttt{km}$ and moves towards $[1,1]\texttt{km}$ at $200\texttt{m/s}$ with a uniform $RCS$ of 100\texttt{$m^2$}. 
These values are consistent with a typical commercial aircraft. 
The radar nodes have access to 8 equally spaced channels of equal bandwidth between 2.4 and 2.5 \texttt{GHz}. 
Each transmitter has an output power of 20\texttt{dBw}, and the arrays have a main beam gain of 30\texttt{dB}. 
The LFM chirp waveforms have a bandwidth of 100\texttt{MHz}. 
Each CPI lasts 10\texttt{ms} and contains 1000 pulses. 
All told, the simulation lasts for 700 CPIs. 
Fig. \ref{fig:Scene} shows one instance of this scene. 

\begin{figure}
    \centering
    \includegraphics[scale=0.65]{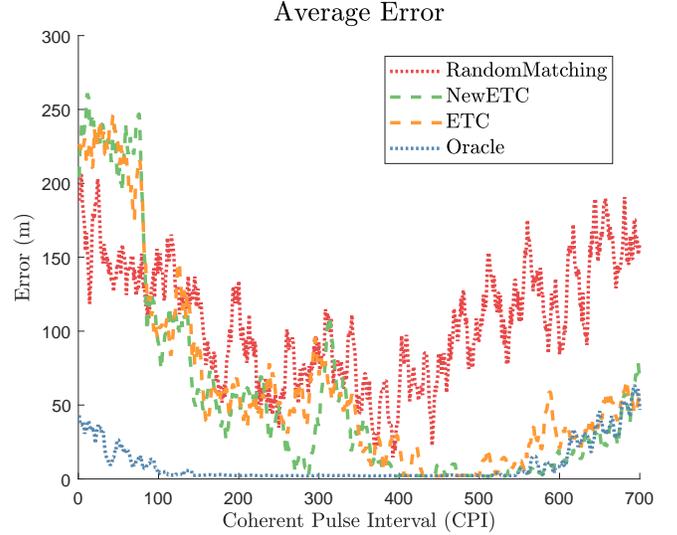}
    \caption{Localization error for the four different algorithms under consideration, averaged over 30 simulations. }
    \label{fig:error}
\end{figure}

We can first examine the error for each algorithm. 
Fig. \ref{fig:error} shows the position tracking error. 
Each CRN uses the observation from all radar nodes to establish a localization estimate once per CPI, which we compare against the target's true location in the middle of the CPI. 
We can see that the oracle exhibits the best performance for the entire simulation, while Random Matchings has quite variable performance. 
Note that both ETC and ETP experience greater in the initial steps than Random Matchings. This is due to the specified exploration sequence, which explores all possible matchings for several steps. 
Also note the increased error at the beginning and end of the track. 
This is due to the target being relatively further away from the nodes during these periods, as well as the poor geometry. 

We compare an emperical CDF for all four algorithms described above. 
The CDF is generated for the entire simulation, as well as for the last 300 CPIs so that we can see the performance of each algorithm after the exploration period has passed. 

\begin{figure}
    \centering
    \includegraphics[scale=0.65]{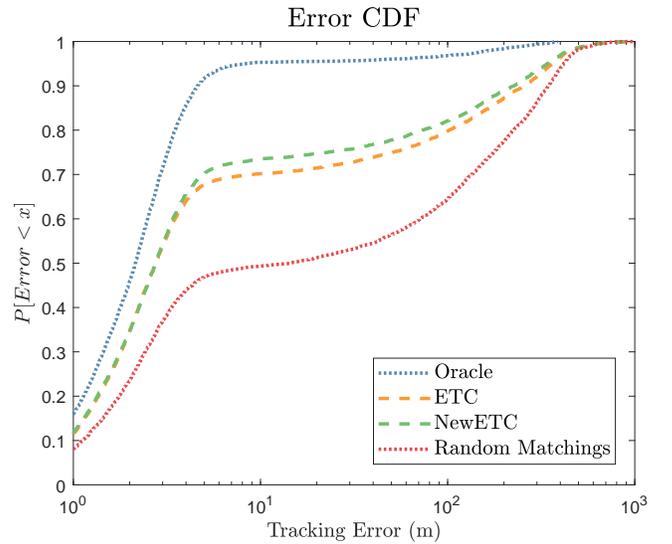}
    \caption{Emperical error CDFs for each algorithm show the relative performance for the entirety of each simulation. As expected, the oracle algorithm obtains the best possible performance given the RF environment. The ETP algorithm is able to exploit knowledge of the target motion to obtain superior results to ETC. Random Matchings provide the lowest performance, since this algorithm does not learn any information throughout the game. }
    \label{fig:ECDF}
\end{figure}

We can expect that the oracle performance will represent an upper bound on realistic performance, while Random Matchings should represent a lower bound. 
In Fig. \ref{fig:ECDF} we see that ETC and ETP performance lies inside this region as expected. 
This empirical CDF represents the probability of a given algorithm obtaining localization error less than a given value. 
ETC and ETP have relatively similar performance, with ETP having a slight advantage. 
Note that there is a substantial gap between Random Matchings and ETC, which demonstrates that any amount of learning is superior to blind action selection.

\begin{figure}
    \centering
    \includegraphics[scale=0.65]{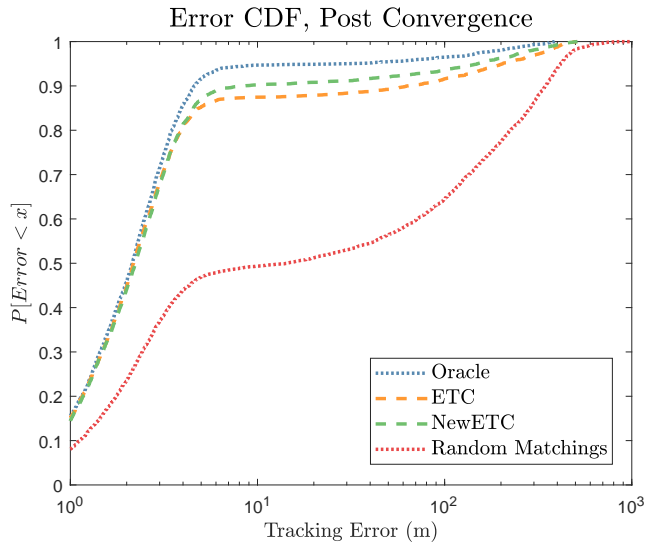}
    \caption{Emperical error CDFs, created using the last 300 CPIs for each simulation. This demonstrates the algorithm \emph{post} convergence, when the algorithm attempts to exploit the information it has learned. There is an even larger gap between ETC and ETP, showing again the benefit of exploiting the target's motion. }
    \label{fig:post_conv}
\end{figure}

In Fig. \ref{fig:post_conv}, we see the empirical CDF for the performance of each algorithm after ETC and ETP converge. 
The performance for the oracle and Random Matchings is the same as in Fig. \ref{fig:ECDF}. 
ETC exhibits good performance in this time period, as it has learned the subset of channels which offer the best \texttt{SINR}. 
ETP, however, has performance which nearly reaches the oracle. 
ETP takes into account the motion of the target, which enables it to predict \texttt{SINR} in the future and take appropriate actions. 

\section{Conclusions}
In this work we have examined a cognitive radar network which is assisted by a central coordinator to select channels over time. 
The network is capable of learning which channels provide better \texttt{SINR}, as well as incorporating target range information to predict future \texttt{SINR} to inform channel selection. 
We detailed online machine learning algorithms this network could use to learn about the environment and provided analysis through simulations.

While previous work has investigated the general CRN learning problem, we contribute techniques which take into account the fact that the target motion will alter the rewards the network will observe, and discuss a method to exploit this information. 
This alteration improves the realism of the problem, as well as changing the applicable algorithms, since the rewards can no longer be assumed to be constant over time. 
We also develop a hybrid network model, where cognition occurs in a central coordinator as well as in each radar node.

The simulations showed that out of those we studied, the proposed ETP algorithm exhibits the best localization performance. 
This is attributed to the ability of ETP to learn the environment and exploiting the target motion to make better selections. 
This performance gap is emphasized when analyzing the performance post convergence, once the algorithm attempts to exploit the information it has learned. 
This is due to the fact that the ETP algorithm is able to utilize information about the target to predict future range and \texttt{SINR}. 
In addition, ETP shows convergence times on the order of 300 time steps, while previous work \cite{howard2021_multiplayerconf} has seen requirements of up to 1000 time steps for convergence under similar reward conditions. 
This comes at the cost of greater computation at each node.

\subsection{Future Work}
This work only considers simple target models. In future work, a class of target RCS models will be introduced, and we will investigate how a radar network should learn which model fits an observed target best. 
In addition, we intend to examine how robust such a network is to loss of nodes or of the central coordinator. Specifically we will determine which algorithms can continue avoiding mutual interference while improving radar tracking performance when a central coordinator has unplanned outages. 

\bibliographystyle{IEEEtran}
\bibliography{bibli}

    
\end{document}